\documentclass[sigconf]{acmart}
\AtBeginDocument{%
  }

\setcopyright{acmlicensed}
\copyrightyear{2018}
\acmYear{2018}
\acmDOI{XXXXXXX.XXXXXXX}
\acmConference[Conference acronym 'XX]{Make sure to enter the correct
  conference title from your rights confirmation email}{June 03--05,
  2018}{Woodstock, NY}
\acmISBN{978-1-4503-XXXX-X/2018/06}

\usepackage{enumitem}
\usepackage{algorithm,algpseudocode,placeins,array,makecell,multirow}

\usepackage{algorithmicx}
\usepackage{algpseudocode}   
\usepackage{amsmath}

\usepackage{enumitem}
\usepackage{lipsum}
\usepackage{bm}

\setcopyright{none}
\renewcommand\footnotetextcopyrightpermission[1]{}
\begin{document}

\title{Learning to Trust: Dynamic Utilization of Retrieval-Augmented Generation for E-commerce Search Relevance}


\author{Tingqiao Xu}
\affiliation{%
  \institution{Shanghai Jiao Tong University}
  \city{Shanghai}
  \country{China}}
\email{phenomenonkj@sjtu.edu.cn}

\author{Shaowei Yao}
\affiliation{%
  \institution{Taobao \& Tmall Group of Alibaba}
  \city{Hangzhou}
  \country{China}}
\email{yaoshaowei@taobao.com}

\author{Chenhe Dong}
\affiliation{%
  \institution{Taobao \& Tmall Group of Alibaba}
  \city{Hangzhou}
  \country{China}}
\email{dongchenhe.dch@alibaba-inc.com}

\author{Yiming Jin}
\affiliation{%
  \institution{Taobao \& Tmall Group of Alibaba}
  \city{Hangzhou}
  \country{China}}
\email{imayking13@gmail.com}

\author{Zerui Huang}
\affiliation{%
  \institution{Taobao \& Tmall Group of Alibaba}
  \city{Hangzhou}
  \country{China}}
\email{huangzerui.hzr@taobao.com}

\author{Dan Ou}
\affiliation{%
  \institution{Taobao \& Tmall Group of Alibaba}
  \city{Hangzhou}
  \country{China}}
\email{oudan.od@taobao.com}

\author{Haihong Tang}
\affiliation{%
  \institution{Taobao \& Tmall Group of Alibaba}
  \city{Hangzhou}
  \country{China}}
\email{piaoxue@taobao.com}

\author{Bo Zheng}
\affiliation{%
  \institution{Taobao \& Tmall Group of Alibaba}
  \city{Beijing}
  \country{China}}
\email{bozheng@alibaba-inc.com}

\renewcommand{\shortauthors}{Trovato et al.}

\begin{abstract}
Accurately estimating query–item relevance is vital for e-commerce ranking and conversion. While Large Language Models (LLMs) excel at reasoning, they often lack specialized knowledge required for long-tail or fast-evolving queries, necessitating Retrieval-Augmented Generation (RAG). However, production environments face three critical challenges: 
(1) external context is inherently noisy and inconsistent; 
(2) extreme latency budgets prohibit multi-stage processing or refinement; and 
(3) the model must simultaneously assess relevance and context-trust within a unified inference pass. 
We propose \textbf{DyKnow-RAG}, a reinforcement learning framework that teaches LLMs to learn to trust through dynamic utilization of external knowledge. Built on Group Relative Policy Optimization (GRPO), DyKnow-RAG utilizes a dual-group rollout strategy (parametric-only vs. with-context) and a posterior-driven inter-group advantage scaling mechanism. This enables the model to optimize context utilization without human process labels or extra inference overhead. Our pipeline further integrates structured Chain-of-Thought (CoT) and an uncertainty-prioritized RL pool to stabilize training.
Offline evaluations show significant Macro-F1 and Accuracy gains, particularly on noise-sensitive query slices. Importantly, DyKnow-RAG has been deployed in Taobao’s production system, serving hundreds of millions of active users and billions of daily search requests. Controlled A/B tests demonstrate consistent lifts in key business metrics, including GSB and Item Goodrate, while maintaining a p99 latency under 400ms. This work provides a scalable and deployable paradigm for operationalizing noisy RAG under extreme efficiency constraints of large-scale industrial search.

\end{abstract}



\received{20 February 2007}
\received[revised]{12 March 2009}
\received[accepted]{5 June 2009}

\maketitle

\section{Introduction}
In large-scale e-commerce search (e.g., Taobao, Amazon), accurate query--item relevance estimation is the cornerstone of ranking quality and user conversion. Over decades, relevance modeling has evolved from hand-crafted statistical scoring like TF-IDF \cite{aizawa2003tf_idf} and BM25 \cite{robertson2009bm25} to deep semantic matching \cite{huang2013dssm, devlin2019bert}. Recently, Large Language Models (LLMs) have reframed relevance as a reasoning task, with frameworks like LREF \cite{tang2025lref}, TaoSR1 \cite{dong2025taosr1}, and ProRBP \cite{chen2024ProRBP} utilizing supervised fine-tuning (SFT) and Direct Preference Optimization (DPO) \cite{rafailov2023dpo} to align LLMs with human judgment. 

Despite these advances, LLMs often struggle with long-tail or knowledge-intensive queries. Retrieval-Augmented Generation (RAG) addresses this by providing external context like community reviews, product evaluations, and attribute encyclopedias. However, operationalizing RAG in production e-commerce faces critical challenges: (1) \textbf{Noise and Heterogeneity:} External context is rife with promotional phrasing, tag clutter, and affective language, requiring robust discrimination. (2) \textbf{Extreme Latency Constraints:} Production e-commerce environments operate under millisecond-level budgets, which strictly prohibit the multi-stage cleaning or iterative reasoning typical of academic RAG frameworks like Self-RAG \cite{asai2024self-rag} or Adaptive-RAG \cite{jeong2024adaptive-rag}. (3) \textbf{Unified Multi-Task Reasoning:} Within this single pass, the model must simultaneously execute two coupled tasks: determining the trustworthiness of the external context and assessing query--item relevance. This requires the model to embed knowledge-trust policies directly into its relevance reasoning logic. While recent RL-based models \cite{jin2025Search-R1, lin2025knowledgeable-r1, song2025R1-Searcher++} explore dynamic knowledge acquisition, adapting them to such highly coupled and time-critical industrial scenarios remains an open challenge.

To address these, we propose \textbf{DyKnow-RAG}, a reinforcement learning framework that teaches LLMs to \textit{learn to trust}. Built on Group Relative Policy Optimization (GRPO) \cite{shao2024grpo}, DyKnow-RAG enables the model to dynamically decide whether to adopt, partially adopt, or ignore retrieved context within a single-pass inference, without auxiliary critique models \cite{asai2024self-rag, zhao2025R-Search} or multi-step retrieval \cite{liu2024RA-ISF, shi2025search_and_refine}. Our approach utilizes a dual-group rollout strategy (parametric-only vs.\ single-context) and introduces a \textit{posterior-driven inter-group advantage scaling} mechanism. This allows the policy to bridge the correctness gap between parametric and non-parametric knowledge without expensive process labels.

Our production-aligned pipeline further optimizes stability through: (i) structured Chain-of-Thought (CoT) that explicitly records context-usage decisions; (ii) an uncertainty-prioritized RL pool from SFT posteriors to focus training on borderline cases; and (iii) an optional DPO warm-start. DyKnow-RAG is deployed in Taobao’s production system, serving hundreds of millions of users and billions of daily requests. Compared to baseline SFT and general RL variants \cite{shao2024grpo, lin2025knowledgeable-r1}, our framework delivers significant offline gains and consistent online A/B lifts while maintaining a p99 latency under 400ms.

Our core contributions are summarized as follows:
\begin{itemize}[leftmargin=*,topsep=0pt,itemsep=2pt]
\item \textbf{Robust and Efficient RAG Integration:} We propose a single-pass framework that integrates noisy external context into e-commerce relevance modeling, effectively managing retrieval inconsistencies within strict production constraints.
\item \textbf{DyKnow-RAG Optimization:} We introduce a posterior-driven inter-group advantage scaling mechanism, enabling label-free dynamic context-utilization tailored for noisy e-commerce data.
\item \textbf{Industrial-Scale Impact:} We demonstrate the first large-scale deployment of dynamic Retrieval-Augmented Generation on Taobao’s massive traffic, achieving consistent business gains and verified reliability.
\end{itemize}

\begin{figure*}[htbp]
    \centering
    \includegraphics[width=0.95\textwidth]{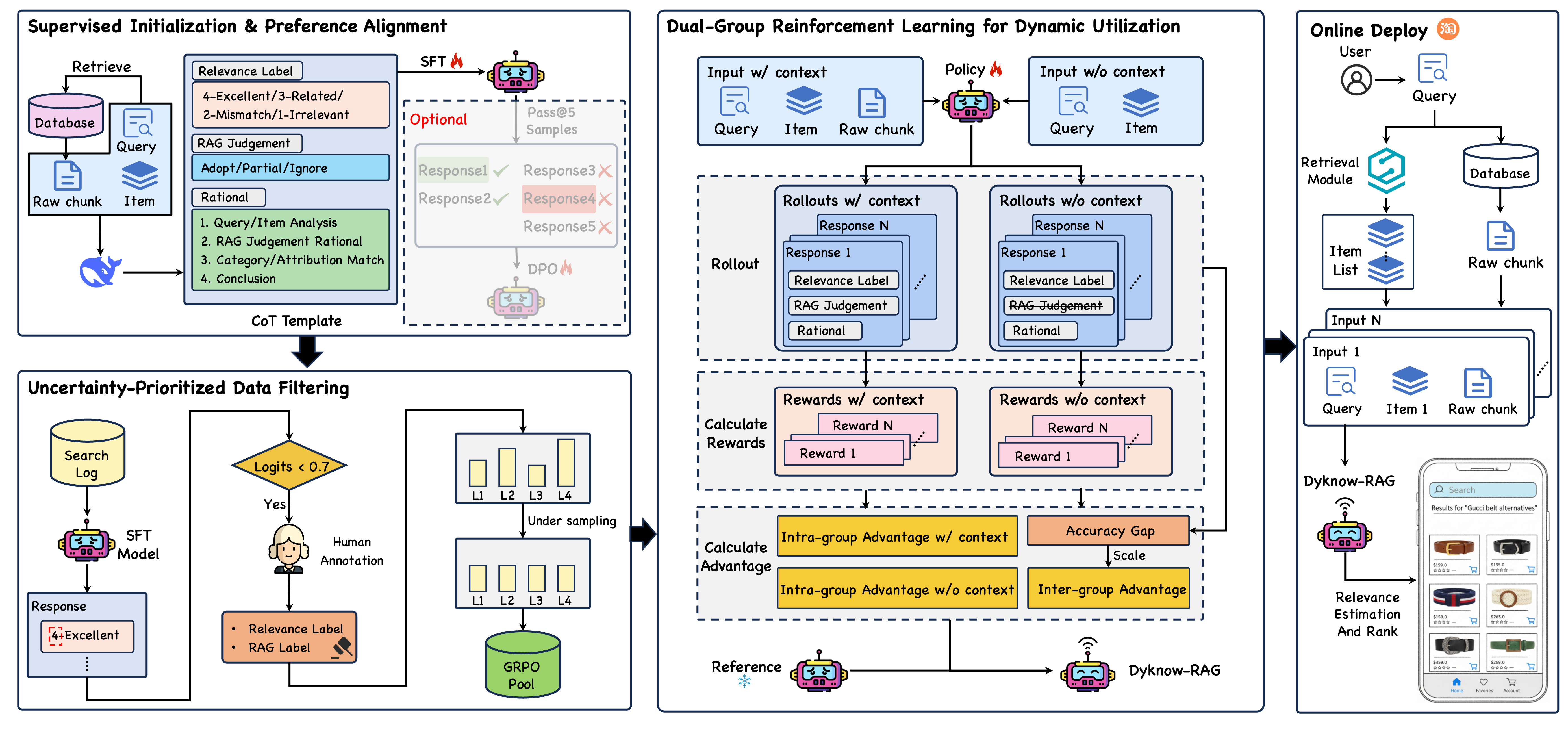}
    \captionsetup{skip=3pt} 
    \caption{Overview of DyKnow-RAG training and deployment workflow.}
    \label{fig:overview}
    \vspace{-4pt} 
\end{figure*}


\section{Method}
DyKnow-RAG is a production-aligned framework designed to intelligently discern when to trust external context. As illustrated in Figure~\ref{fig:overview}, the pipeline comprises three stages: (i) supervised initialization and preference alignment; (ii) uncertainty-prioritized data filtering; and (iii) dual-group reinforcement learning for dynamic knowledge utilization.
\subsection{Task Formulation and Reward Signal}
We formalize e-commerce relevance as a classification task. Given a query $q$ and an item $i$, we retrieve the most relevant context chunk $c^*$. The input is denoted as $x = (q, i, c^*)$. The model is required to generate a response $o$ that includes a four-tier relevance label $\hat{y} \in \{L_4, L_3, L_2, L_1\}$ and a supporting rationale.

\noindent\textbf{Binary Correctness Reward:}\quad To ensure stability in noisy e-commerce environments, we utilize a \textbf{binary correctness-based reward} rather than a dense reward model, which mitigates reward hacking. For a sampled response $o_j$, the reward $R(o_j)$ is defined as:

\begin{equation}
R(o_j) = 
\begin{cases} 
1, & \text{if } \text{parse}(\hat{y}) = y_{\text{gold}} \\
0, & \text{otherwise}
\end{cases}
\end{equation}

where $\text{parse}(\cdot)$ extracts the relevance tier from the structured output and $y_{\text{gold}}$ is the human-annotated ground truth. This binary signal ensures that policy optimization is strictly aligned with the ultimate objective: classification accuracy.

\subsection{Policy Initialization and Data Filtering}
\noindent\textbf{Stage 1: SFT with Structured CoT.}\quad We initialize the model using a structured Chain-of-Thought (CoT) template: \texttt{[Relevance Tier] + [Trust Decision] + [Rationale]}. The "Trust Decision" step explicitly requires the model to judge the utility of $c^*$. This supervised stage provides a critical warm-start.

\noindent\textbf{Stage 2: Preference Alignment (Optional DPO).}\quad To further calibrate the policy before RL, we optionally perform Direct Preference Optimization (DPO) \cite{rafailov2023dpo}. For each instance, the SFT model generates five samples (pass@5). We construct preference pairs $(y^+, y^-)$ where $y^+$ represents correct relevance judgments and $y^-$ represents incorrect ones. This stage ensures the model prioritizes correct reasoning paths, establishing a more accurate and stable policy before the RL phase.

\noindent\textbf{Stage 3: Uncertainty-Prioritized Filtering.}\quad To maximize RL sample efficiency, we prioritize "borderline" cases. We calculate the posterior probability $p_\theta(\hat{t} | x)$ for the SFT model's predicted relevance token. Instances with $p_\theta < 0.7$ (empirically corresponding to a $\approx 50\%$ error rate) are routed into the RL pool. This ensures DyKnow-RAG concentrates on contentious cases where the model struggles to reconcile parametric knowledge with external evidence, providing a strong learning signal to refine its trust policy.

\subsection{Reinforcement Learning for Dynamic Utilization}
We employ \textbf{Group Relative Policy Optimization (GRPO)} \cite{shao2024grpo} to update the policy via group-relative advantages, eliminating the need for a complex value network.

\subsubsection{Dual-Group Rollout Strategy}
Per instance, we generate two distinct groups of $\bm{n}$ rollouts to expose the performance gap between \textbf{parametric-only reasoning} and \textbf{with-context reasoning}:
\begin{itemize}[leftmargin=*,noitemsep]
    \item \textbf{Group 0 (Parametric-only):} Sampled under prompt $p$, containing only $(q, i)$.
    \item \textbf{Group 1 (With-context):} Sampled under prompt $p'$, containing $(q, i, c^*)$.
\end{itemize}
Let $R^{(g)}_j$ be the return of the $j$-th rollout in group $g \in \{0, 1\}$. We compute group-specific statistics to normalize advantages within each group:
\begin{equation}
\mu_g = \frac{1}{n}\sum_{j=1}^{n} R^{(g)}_{j}, \quad s_g = \sqrt{\frac{1}{n}\sum_{j=1}^{n}(R^{(g)}_{j}-\mu_g)^2+\varepsilon}
\end{equation}
The intra-group advantage is $A^{(g)}_{j} = (R^{(g)}_{j} - \mu_g) / s_g$.

\subsubsection{Posterior-Driven Inter-Group Scaling}
To teach the model to "learn to trust," we introduce an inter-group alignment mechanism. We first compute union statistics across both groups:
\begin{equation}
\mu_\star = \frac{\sum R^{(0)}_j + \sum R^{(1)}_j}{2n}, \quad s_\star = \sqrt{\frac{\sum (R^{(g)}_j - \mu_\star)^2}{2n} + \varepsilon}
\end{equation}
The inter-group advantage for Group 0 is $\tilde{A}_j = (R^{(0)}_j - \mu_\star) / s_\star$. We then apply a piecewise scaling function $T(\tilde{A}_j)$ based on the batch-level accuracy gap:
\begin{equation}
T(\tilde{A}_j) = \begin{cases} \alpha \tilde{A}_j, & \text{if } \tilde{A}_j > 0 \\ \beta \tilde{A}_j, & \text{if } \tilde{A}_j \le 0 \end{cases}
\end{equation}

where $\beta = 4 \sigma(4(\text{acc}_{\text{with}} - \text{acc}_{\text{para}}))$ and $\alpha = 0.1 / \beta$. Here, $\text{acc}_{\text{with}}$ and $\text{acc}_{\text{para}}$ denote the average rewards of the with-context and parametric-only groups, respectively. This mechanism forces the policy to prioritize external evidence when it provides a clear correctness gain over internal parametric knowledge.

\subsubsection{Policy Optimization}
The total objective $\mathcal{J}(\theta)$ combines the standard GRPO losses ($\ell^{(0)}, \ell^{(1)}$) with an inter-group alignment term $\hat{\ell}(\theta)$:
\vspace{-9pt}

\begin{equation}
\begin{aligned}
\ell^{(0)}(\theta)
&= \frac{1}{n}\sum_{j=1}^{n}\frac{1}{|o^{(0)}_j|}\sum_{t=1}^{|o^{(0)}_j|}
\min\!\Biggl[
\frac{\pi_{\theta}\!\left(o^{(0)}_{j,t}\mid p,o^{(0)}_{j,<t}\right)}
{\pi_{\theta_{\mathrm{old}}}\!\left(o^{(0)}_{j,t}\mid p,o^{(0)}_{j,<t}\right)}A^{(0)}_j,\\
&\qquad\qquad
\mathrm{clip}\Biggl(
\frac{\pi_{\theta}\!\left(o^{(0)}_{j,t}\mid p,o^{(0)}_{j,<t}\right)}
{\pi_{\theta_{\mathrm{old}}}\!\left(o^{(0)}_{j,t}\mid p,o^{(0)}_{j,<t}\right)};
1-\epsilon,1+\epsilon
\Biggr)A^{(0)}_j
\Biggr].
\end{aligned}
\end{equation}

\begin{equation}
\begin{aligned}
\ell^{(1)}(\theta)
&= \frac{1}{n}\sum_{j=1}^{n}\frac{1}{|o^{(1)}_j|}\sum_{t=1}^{|o^{(1)}_j|}
\min\!\Biggl[
\frac{\pi_{\theta}\!\left(o^{(1)}_{j,t}\mid p,o^{(1)}_{j,<t}\right)}
{\pi_{\theta_{\mathrm{old}}}\!\left(o^{(1)}_{j,t}\mid p,o^{(1)}_{j,<t}\right)}A^{(1)}_j,\\
&\qquad\qquad
\mathrm{clip}\Biggl(
\frac{\pi_{\theta}\!\left(o^{(1)}_{j,t}\mid p,o^{(1)}_{j,<t}\right)}
{\pi_{\theta_{\mathrm{old}}}\!\left(o^{(1)}_{j,t}\mid p,o^{(1)}_{j,<t}\right)};
1-\epsilon,1+\epsilon
\Biggr)A^{(1)}_j
\Biggr].
\end{aligned}
\end{equation}

The alignment term $\hat{\ell}(\theta)$ evaluates parametric-only sequences $o^{(0)}$ under the with-context prompt $p'$:
\begin{equation}
\hat{\ell}(\theta) = \frac{1}{n}\sum_{j=1}^{n}\frac{1}{|o^{(0)}_j|}\sum_{t=1}^{|o^{(0)}_j|} \left[ \pi_\theta(o^{(0)}_{j,t} \mid p', o^{(0)}_{j,<t}) T(\tilde{A}_j) \right]
\end{equation}
The final objective $\mathcal{J}(\theta) = \ell^{(0)}(\theta) + \ell^{(1)}(\theta) + \hat{\ell}(\theta) - \lambda_{\text{KL}} \mathbb{D}_{\text{KL}}[\pi_\theta \mid \mid \pi_{\text{ref}}]$ encourages the policy to autonomously arbitrate between parametric knowledge and external context based on which provides a more reliable correctness signal.

\section{Experiments}
We evaluate DyKnow-RAG through rigorous offline benchmarks and large-scale online A/B tests on Taobao.

\subsection{Experimental Setup}
\noindent\textbf{Expert-Annotated Gold Benchmark:}\quad 
While industrial logs are vast, high-precision labels are scarce. We curated a high-fidelity \textbf{Gold Standard Benchmark} expert-annotated into four tiers: $L_4$ (Excellent, 60\%), $L_3$ (Related, 7\%), $L_2$ (Mismatch, 27\%), and $L_1$ (Irrelevant, 6\%). To stress-test the model's ability to "learn to trust" under noise, the benchmark focuses on four knowledge-intensive categories: (1) \textbf{Q\&A} (interrogative queries), (2) \textbf{Knowledge} (domain-specific expertise), (3) \textbf{Negation} (containing negative terms), and (4) \textbf{Alternatives} (seeking substitute products). These slices represent the most noise-sensitive scenarios in e-commerce search, providing a rigorous testbed for dynamic context utilization.

\noindent\textbf{Implementation Details:}\quad 
Our base model TBStar is fine-tuned for 2 epochs (LR $1\!\times\!10^{-6}$, batch 1024). GRPO RL uses a rollout group size $G{=}16$, KL penalty $\beta{=}0.02$, and uncertainty filtering threshold $p_\theta < 0.7$. Training is implemented on the ROLL framework \cite{wang2025roll}.

\noindent\textbf{Evaluation Metrics:}\quad 
Offline performance is measured by \textbf{Macro F1} and \textbf{Accuracy}. Online evaluation utilizes \textbf{GSB} (human side-by-side comparison) and \textbf{Goodrate} (absolute relevance).

\subsection{Offline Performance}
We evaluate DyKnow-RAG against strong baselines, including parametric-only SFT and various RAG-augmented variants. As shown in Table~\ref{tab:overall_performance}, DyKnow-RAG (DPO-based) achieves the best overall performance, particularly in Macro F1, which reflects a more balanced trust policy.

\noindent\textbf{Key Findings:}\quad 
(1) \textbf{Blind Trust Penalty:} Direct RAG-SFT slightly degrades Macro F1 compared to SFT-only. This suggests that without a dynamic utilization policy, the model "blindly trusts" noisy external context, causing confusion in boundary classes. 
(2) \textbf{Adaptive Trusting:} DyKnow-RAG significantly outperforms vanilla RAG-GRPO, especially in $L_1$ and $L_3$ tiers (C1/C3). This confirms that our posterior-driven scaling effectively teaches the model to autonomously arbitrate between parametric knowledge and external context. 
(3) \textbf{Alignment Synergy:} Initializing from a DPO-calibrated base yields better overall performance than the SFT initialization, further enhancing Accuracy as the model benefits from both improved preference alignment and robust context-gating.

\vspace{-6pt}
\begin{table}[h]
  \caption{Offline Performance on Gold Benchmark}
  \vspace{-6pt}
  \label{tab:overall_performance}
  \small
  \centering
  \setlength{\tabcolsep}{2pt} 
  \begin{tabular}{lcccccc}
    \toprule
    \textbf{Models} & \textbf{C1-F1} & \textbf{C2-F1} & \textbf{C3-F1} & \textbf{C4-F1} & \textbf{Macro-F1} & \textbf{Acc}\\
    \midrule
    SFT-only            & 41.97 & 63.35 & 39.44 & 79.55 & 58.14 & 71.88   \\
    RAG SFT             & 40.47 & 62.59 & 39.21 & 80.37 & 57.06 & 71.92  \\ 
    RAG DPO             & 45.66 & 65.07 & 44.26 & 83.54 & 59.67 & 74.91  \\ 
    RAG GRPO            & 47.54 & 62.72 & 42.82 & 80.18 & 58.33 & 72.55  \\
    DyKnow-RAG (SFT)    & \textbf{49.48} & 64.10 & \textbf{45.64} & 82.02 & 60.33 & 73.26  \\ 
    DyKnow-RAG (DPO)    & 47.96 & \textbf{65.21} & 44.98 & \textbf{83.66} & \textbf{60.45} & \textbf{75.19}  \\ 
    \bottomrule
  \end{tabular}
  \vspace{-4pt}
\end{table}
\vspace{-6pt}

\subsection{Ablation Study}
To isolate the impact of our posterior-driven inter-group scaling, we compare DyKnow-RAG against heuristic and supervised gating strategies (Table~\ref{tab:ablation_gating_vs_dyknow}).

\noindent\textbf{Superiority of Learned Trust:}\quad 
While \textit{Fixed Gating} (up-weighting parametric-only paths) reduces misuse, it fails to utilize context when it is actually beneficial. \textit{Human Gating} (using labels to guide scaling) improves results but is limited by the annotators' prior knowledge, which may not align with the model's internal gaps. DyKnow-RAG outperforms both, demonstrating that posterior-driven RL discovers more nuanced trust policies by directly optimizing the correctness gap.

\noindent\textbf{Robustness at Inference:}\quad 
We also test DyKnow-RAG's performance when the context chunk is intentionally removed at inference. The model maintains high accuracy (74.80\%) in parametric-only mode, proving it has learned a robust fallback policy rather than becoming dependent on external cues.

\vspace{-6pt}
\begin{table}[h]
  \vspace{-2pt}
  \caption{Ablation: Gating Strategy and Inference Mode}
  \label{tab:ablation_gating_vs_dyknow}
  \vspace{-6pt}
  \small
  \centering
  \setlength{\tabcolsep}{4pt}
  \begin{tabular}{lcc}
    \toprule
    \textbf{Setting} & \textbf{Macro F1} & \textbf{Accuracy} \\
    \midrule
    Fixed Gating ($\alpha{=}2, \beta{=}0.05$) & 59.31 & 71.75 \\
    Human Context-use Gating & 60.15 & 73.05 \\
    DyKnow-RAG (Inference w/ context) & \textbf{60.45} & \textbf{75.19} \\
    DyKnow-RAG (Inference w/o context) & 59.71 & 74.80 \\
    \bottomrule
  \end{tabular}
  \vspace{-6pt}
\end{table}
\vspace{-6pt}

\subsection{Online Impact and System Deployment}
DyKnow-RAG has been successfully deployed in Taobao's production stack since September 2025. We conducted a 7-day controlled A/B test on 4\% traffic over eligible queries, involving hundreds of millions of search requests to verify its industrial-scale reliability.

\noindent\textbf{Relevance Quality:}\quad 
We selected 2,000 queries, requested both experimental buckets for each query,
and compared the top-10 results across three metrics. Table~\ref{tab:online_eval_table} shows DyKnow-RAG achieves substantial lifts in human-evaluated GSB, particularly on Q\&A (+10.37\%) and Knowledge (+7.75\%) queries. These gains confirm that the model effectively utilizes external context to resolve complex intents. The positive Item Goodrate across all slices further validates that our trust policy reduces noise-induced misjudgments.

\noindent\textbf{Business Metrics:}\quad 
Global traffic metrics show a \textbf{+0.1\% GMV} lift and significant increases in user engagement, with \textbf{IPV +2.35\%} (Item Page Views) and \textbf{PV +2.13\%}. While Orders showed a slight fluctuation (-0.5\%), the overall engagement growth indicates that users are finding higher-quality, relevant items to browse.

\noindent\textbf{System Efficiency:}\quad 
Operating under Taobao's extreme concurrency, DyKnow-RAG maintains a \textbf{p99 latency < 400ms}. The \textbf{MFU (Model FLOPs Utilization)} is approximately 35\%, demonstrating that our single-pass architecture delivers RAG-augmented reasoning with no extra inference overhead compared to vanilla LLMs.

\vspace{-4pt}
\begin{table}[h]
  \caption{Online A/B Test Results (Billions of Requests)}
  \vspace{-6pt}
  \label{tab:online_eval_table}
  \small
  \centering
  \setlength{\tabcolsep}{2.5pt}
  \begin{tabular}{lccc}
    \toprule
    \textbf{Query Slice} & \textbf{GSB Rel. Gain} & \textbf{Query Goodrate} & \textbf{Item Goodrate} \\
    \midrule
    Q\&A      & +10.37\% & +5.76 pt & +2.45 pt \\
    Knowledge & +7.75\%  & +2.91 pt & +1.15 pt \\
    Negation  & +3.25\%  & -0.20 pt & +0.81 pt \\
    Alternatives & +3.60\%  & +1.80 pt & +0.84 pt \\
    \bottomrule
  \end{tabular}
  \vspace{-4pt}
\end{table}
\vspace{-7pt}

\subsection{Case Study}

\begin{figure}[h]
  \centering
  \includegraphics[width=0.95\linewidth]{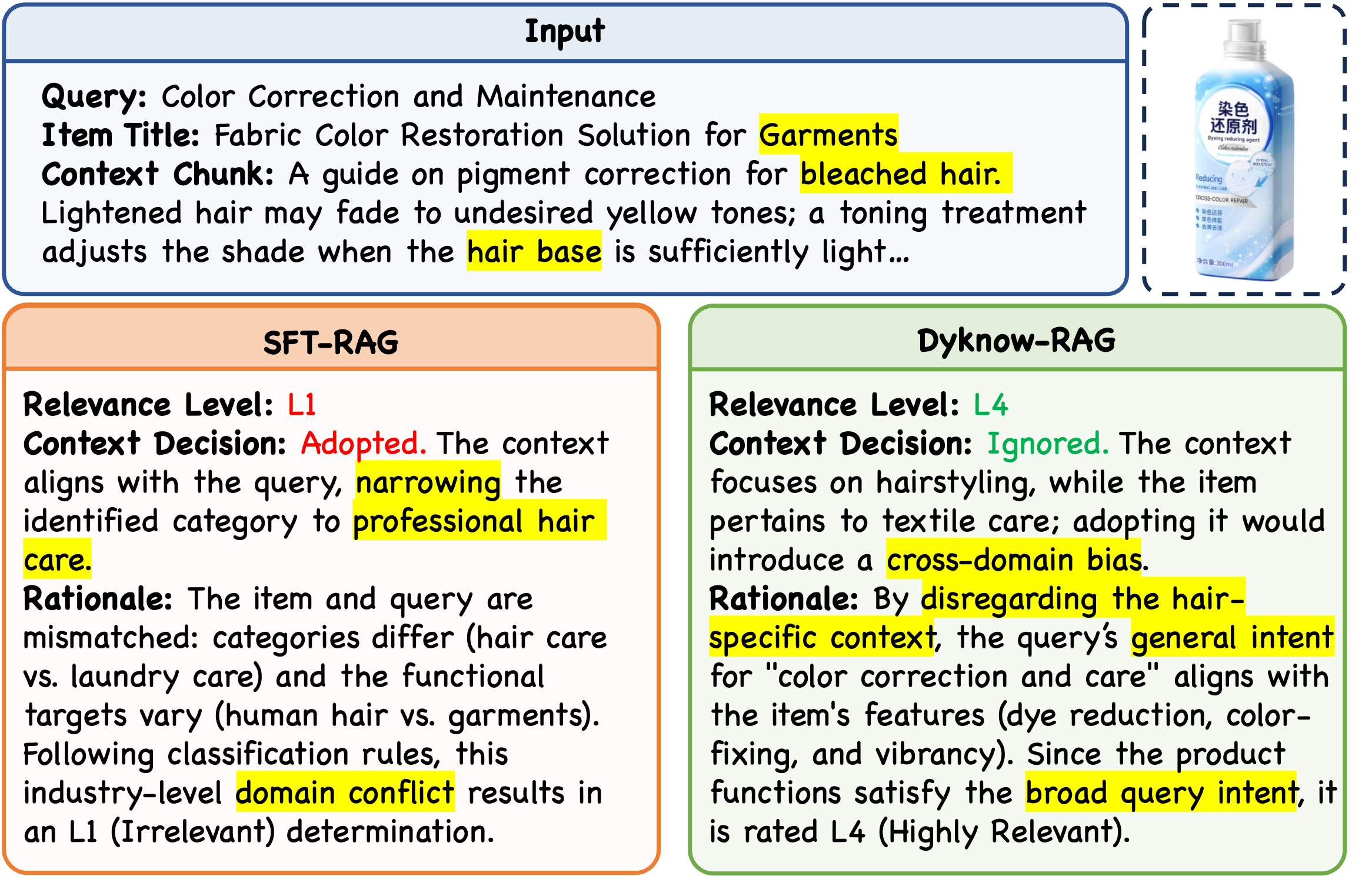}
  \caption{Case study of DyKnow-RAG robustness against noisy context.}
  \label{fig:case_study}
\end{figure}

Figure~\ref{fig:case_study} showcases DyKnow-RAG’s robustness. Unlike Vanilla RAG-SFT, which "blindly trusts" misleading hair-care context (L1), DyKnow-RAG identifies the discrepancy and relies on parametric knowledge to deliver the correct L4 rating, effectively resisting intent distortion.
\vspace{-4pt}

\vspace{-13pt}

\section{Conclusion}
We proposed \textbf{DyKnow-RAG}, a single-pass reinforcement learning framework that enables LLMs to "learn to trust" noisy external context in e-commerce search. By utilizing dual-group GRPO with posterior-driven advantage scaling, the model autonomously arbitrates between internal parametric knowledge and inconsistent retrieval signals. Our large-scale deployment at Taobao proves that DyKnow-RAG delivers consistent improvements in both \textbf{Relevance Quality} and \textbf{Business Efficiency}. While serving billions of daily search requests, the system maintains a sub-400ms latency without extra inference overhead. This provides a scalable and robust paradigm for operationalizing Retrieval-Augmented Generation in high-concurrency production environments.


\newpage\section*{Presenter Bio}
Tingqiao Xu is an M.Sc. candidate in the Department of Automation at Shanghai Jiao Tong University. His research focuses on Large Language Models for search systems, specifically in the areas of Retrieval-Augmented Generation (RAG) and Reinforcement Learning. His work aims to address challenges in dynamic knowledge utilization and reasoning reliability within large-scale search environments.

\section*{Presentation Preference}
Poster

\appendix
\bibliographystyle{ACM-Reference-Format}  
\bibliography{refs}

@article{asai2024self-rag,
  title={Self-rag: Learning to retrieve, generate, and critique through self-reflection},
  author={Asai, Akari and Wu, Zeqiu and Wang, Yizhong and Sil, Avirup and Hajishirzi, Hannaneh},
  year={2024},
  publisher={ICLR}
}

@inproceedings{jeong2024adaptive-rag,
  title={Adaptive-RAG: Learning to Adapt Retrieval-Augmented Large Language Models through Question Complexity},
  author={Jeong, Soyeong and Baek, Jinheon and Cho, Sukmin and Hwang, Sung Ju and Park, Jong C},
  booktitle={Proceedings of the 2024 Conference of the North American Chapter of the Association for Computational Linguistics: Human Language Technologies (Volume 1: Long Papers)},
  pages={7029--7043},
  year={2024}
}

@article{song2025R1-Searcher++,
  title={R1-Searcher++: Incentivizing the Dynamic Knowledge Acquisition of LLMs via Reinforcement Learning},
  author={Song, Huatong and Jiang, Jinhao and Tian, Wenqing and Chen, Zhipeng and Wu, Yuhuan and Zhao, Jiahao and Min, Yingqian and Zhao, Wayne Xin and Fang, Lei and Wen, Ji-Rong},
  journal={arXiv preprint arXiv:2505.17005},
  year={2025}
}

@inproceedings{liu2024RA-ISF,
  title={RA-ISF: Learning to Answer and Understand from Retrieval Augmentation via Iterative Self-Feedback},
  author={Liu, Yanming and Peng, Xinyue and Zhang, Xuhong and Liu, Weihao and Yin, Jianwei and Cao, Jiannan and Du, Tianyu},
  booktitle={Findings of the Association for Computational Linguistics ACL 2024},
  pages={4730--4749},
  year={2024}
}

@article{jin2025Search-R1,
  title={Search-r1: Training llms to reason and leverage search engines with reinforcement learning},
  author={Jin, Bowen and Zeng, Hansi and Yue, Zhenrui and Yoon, Jinsung and Arik, Sercan and Wang, Dong and Zamani, Hamed and Han, Jiawei},
  journal={arXiv preprint arXiv:2503.09516},
  year={2025}
}

@article{shi2025search_and_refine,
  title={Search and Refine During Think: Autonomous Retrieval-Augmented Reasoning of LLMs},
  author={Shi, Yaorui and Li, Sihang and Wu, Chang and Liu, Zhiyuan and Fang, Junfeng and Cai, Hengxing and Zhang, An and Wang, Xiang},
  journal={arXiv preprint arXiv:2505.11277},
  year={2025}
}

@article{zhao2025R-Search,
  title={R-Search: Empowering LLM Reasoning with Search via Multi-Reward Reinforcement Learning},
  author={Zhao, Qingfei and Wang, Ruobing and Xu, Dingling and Zha, Daren and Liu, Limin},
  journal={arXiv preprint arXiv:2506.04185},
  year={2025}
}

@article{lin2025knowledgeable-r1,
  title={Knowledgeable-r1: Policy Optimization for Knowledge Exploration in Retrieval-Augmented Generation},
  author={Lin, Chenyu and Wen, Yilin and Su, Du and Sun, Fei and Chen, Muhan and Bao, Chenfu and Lv, Zhonghou},
  journal={arXiv preprint arXiv:2506.05154},
  year={2025}
}

@article{robertson2009bm25,
  title={The probabilistic relevance framework: BM25 and beyond},
  author={Robertson, Stephen and Zaragoza, Hugo and others},
  journal={Foundations and Trends{\textregistered} in Information Retrieval},
  volume={3},
  number={4},
  pages={333--389},
  year={2009},
  publisher={Now Publishers, Inc.}
}

@article{aizawa2003tf_idf,
  title={An information-theoretic perspective of tf--idf measures},
  author={Aizawa, Akiko},
  journal={Information Processing \& Management},
  volume={39},
  number={1},
  pages={45--65},
  year={2003},
  publisher={Elsevier}
}

@inproceedings{devlin2019bert,
  title={Bert: Pre-training of deep bidirectional transformers for language understanding},
  author={Devlin, Jacob and Chang, Ming-Wei and Lee, Kenton and Toutanova, Kristina},
  booktitle={Proceedings of the 2019 conference of the North American chapter of the association for computational linguistics: human language technologies, volume 1 (long and short papers)},
  pages={4171--4186},
  year={2019}
}

@inproceedings{tang2025lref,
  title={LREF: A Novel LLM-based Relevance Framework for E-commerce Search},
  author={Tang, Tian and Tian, Zhixing and Zhu, Zhenyu and Wang, Chenyang and Hu, Haiqing and Tang, Guoyu and Liu, Lin and Xu, Sulong},
  booktitle={Companion Proceedings of the ACM on Web Conference 2025},
  pages={468--475},
  year={2025}
}

@article{dong2025taosr1,
  title={TaoSR1: The Thinking Model for E-commerce Relevance Search},
  author={Dong, Chenhe and Yao, Shaowei and Jiao, Pengkun and Yang, Jianhui and Jin, Yiming and Huang, Zerui and Zhou, Xiaojiang and Ou, Dan and Tang, Haihong},
  journal={arXiv preprint arXiv:2508.12365},
  year={2025}
}

@article{chen2024ProRBP,
  title={Towards Boosting LLMs-driven Relevance Modeling with Progressive Retrieved Behavior-augmented Prompting},
  author={Chen, Zeyuan and Wu, Haiyan and Wu, Kaixin and Chen, Wei and Zhong, Mingjie and Xu, Jia and Liu, Zhongyi and Zhang, Wei},
  journal={arXiv preprint arXiv:2408.09439},
  year={2024}
}

@article{rafailov2023dpo,
  title={Direct preference optimization: Your language model is secretly a reward model},
  author={Rafailov, Rafael and Sharma, Archit and Mitchell, Eric and Manning, Christopher D and Ermon, Stefano and Finn, Chelsea},
  journal={Advances in neural information processing systems},
  volume={36},
  pages={53728--53741},
  year={2023}
}

@article{shao2024grpo,
  title={Deepseekmath: Pushing the limits of mathematical reasoning in open language models},
  author={Shao, Zhihong and Wang, Peiyi and Zhu, Qihao and Xu, Runxin and Song, Junxiao and Bi, Xiao and Zhang, Haowei and Zhang, Mingchuan and Li, YK and Wu, Yang and others},
  journal={arXiv preprint arXiv:2402.03300},
  year={2024}
}

@inproceedings{huang2013dssm,
  title={Learning deep structured semantic models for web search using clickthrough data},
  author={Huang, Po-Sen and He, Xiaodong and Gao, Jianfeng and Deng, Li and Acero, Alex and Heck, Larry},
  booktitle={Proceedings of the 22nd ACM international conference on Information \& Knowledge Management},
  pages={2333--2338},
  year={2013}
}

@article{wang2025roll,
  title={Reinforcement Learning Optimization for Large-Scale Learning: An Efficient and User-Friendly Scaling Library},
  author={Wang, Weixun and Xiong, Shaopan and Chen, Gengru and Gao, Wei and Guo, Sheng and He, Yancheng and Huang, Ju and Liu, Jiaheng and Li, Zhendong and Li, Xiaoyang and others},
  journal={arXiv preprint arXiv:2506.06122},
  year={2025}
}
\end{document}